\documentclass{ptephy_v1}%%%%%% to generate preprint number with ptep logo

\preprintnumber{XXXX-XXXX} %%% %%% Insert preprint number here

\usepackage{graphics}% for dealing with figures.

\usepackage{subfigure} %for getting the subfigures e.g., "Figure 1a and 1b" etc.

\usepackage{color}

\newcommand{\Am}{$^{241}$Am}

\newcommand{\Cs}{$^{137}$Cs}

\newcommand{\Ba}{$^{133}$Ba}

\newcommand{\Cd}{$^{109}$Cd}

\newcommand{\zwo}{ZnWO$_4$ }

\begin{document}

\title{Anisotropic Response Measurements of \zwo Scintillators to Neutrons for Developing the Direction-Sensitive Dark Matter Detector}

%%%% To generate auto affiliation numbers please use \author{}\affil{} command

\author[1,*]{Juan W. Pedersen}

\author[2,3]{Hiroyuki Sekiya}

\author[2,3]{Koichi Ichimura}

\affil[1]{Graduate School of Arts and Sciences, University of Tokyo, Komaba, Meguro-ku, Tokyo, 153-8902, Japan}

\affil[2]{Kamioka Observatory, Institute for Cosmic Ray Research, University of Tokyo, Hida, Gifu, 506-1205, Japan}

\affil[3]{Kavli Institute for the Physics and Mathematics of the Universe (WPI), the University of Tokyo, Kashiwa, Chiba, 277-8582, Japan}

\affil[*]{{\rm E-mail: pedersen@hep1.c.u-tokyo.ac.jp}}

%%\affil[$\dag$]{{\rm E-mail: sekiya@icrr.u-tokyo.ac.jp}}

%%\affil[$\ddag$]{{\rm E-mail: ichimura@km.icrr.u-tokyo.ac.jp}}

%\thanks{These authors contributed equally to this work}}

%%% To include the collaborator name... Please use the command "\collaborator"

%%% For example: \collaborator{ATLAS Collaboration}

\begin{abstract}%
The scintillation yields of \zwo crystals change depending on the incident direction of particles. This property can be used as a direction-sensitive dark matter detector. So, we investigated the \zwo light yields ratio of neutron-induced nuclear recoils to gamma-ray induced electron recoils(quenching factor). \par
Two surfaces almost perpendicular to the crystal axis of \zwo were irradiated with a quasi-monochromatic neutron beam of 0.885MeV, and the quenching factors of both surfaces for the oxygen-nucleus recoil in the \zwo crystal were measured. The obtained quenching factors of the two surfaces were 0.235 $\pm$ 0.026 and 0.199 $\pm$ 0.020, respectively confirming 15.3\% of anisotropy.
\end{abstract}

\subjectindex{C34, C40,C43}

\maketitle

\section{Introduction}

Galactic halos are thought to be composed of weak interacting massive particles (WIMP). These particles are directly detectable by measuring the nuclear recoils associated with elastic scattering. However, to distinguish between WIMP and background scattering events in normal detectors, the annual modulation of the event rate due to revolution of the Earth must be detected\cite{zero}. \par
Alternatively, if the incident direction of WIMP is measured, then accurate results of WIMP detection can be provided\cite{one}. Currently, many gaseous particle track detectors are being developed for direction-sensitive WIMP detection\cite{two}. However, it is also important to obtain target mass to search for small cross-sectional events. Accordingly, solid direction-sensitive detectors have also been developed. For example, the scintillation emission efficiency of organic crystal scintillators for heavily charged particles changes depending on the incident direction of the particles. \par
A WIMP detector using this property has been previously investigated\cite{sekiya}. WIMP detectors detect the diurnal modulation spectrum of the scintillation emission due to variations in the incident direction of WIMP with respect to the crystal associated with the Earth rotation. Because the target of organic scintillators is limited to light nuclei such as hydrogen and carbon, they are not suitable for some prevailing candidates for WIMP such as pure WINO and other heavy SUSY particles\cite{hisano}.\par
In 2013, the dependence of scintillation yields on the incident direction of $ \alpha $ particles was reported in \zwo crystal, which is an inorganic scintillator containing heavy nuclei\cite{adamo,fedor}. In this paper, we evaluated the anisotropic scintillation response of a \zwo crystal by nuclear recoil using neutrons to examine its performance as a direction-sensitive dark matter detector.

\section{Setup}

\subsection{\zwo crystal}

\begin{figure}[!h]

\centering

\includegraphics[width=3.0in]{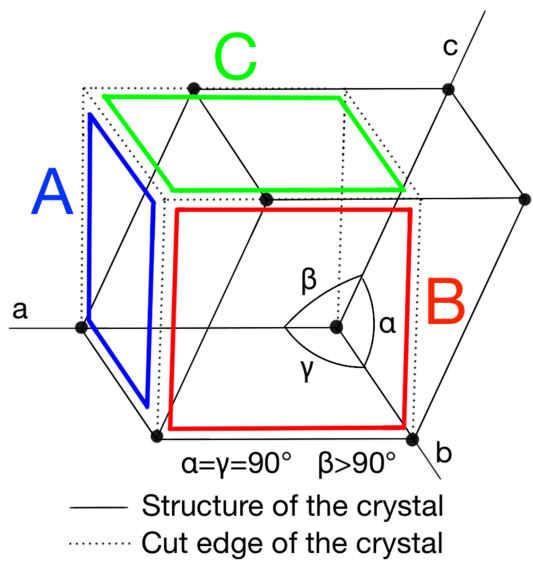}

\caption{The structure of a unit cell of \zwo crystals}

\label{fig:crystal}

\end{figure}

The properties of the \zwo crystals are given in Table \ref{tb:info}, and the information regarding length and angle of the basic vector of the unit cell is given in Table \ref{tb:unitcell}. The \zwo crystals are monoclinic crystals. Two sets of three basic vectors of the unit cell intersect at the right angle, and the other pair has an obtuse angle. A schematic diagram of the crystal structure is given in Fig. \ref{fig:crystal}. \par

In this study, a cubic crystal with a size of 2 cm$\times$2 cm$\times$2 cm was used, as shown in Fig. \ref{fig:crystal}. Although this crystal is monoclinic, its shape is that of a rectangular parallelepiped as the obtuse angle is $ \beta = 90.62^\circ $. We cut the crystal to match the shape of a unit cell. Accordingly, it has the same bottom surface as the unit cell and the side surfaces are parallel to the unit cell. \par

In Fig. \ref{fig:crystal}, the A, B, and C surfaces correspond to the surfaces perpendicular to a, b, and c axis, respectively. Each back side is defined as A$^\prime$, B$^\prime$, and C$^\prime$ surface. The B surface is perpendicular to the optical axis (b-axis) of the crystal. The structure of the crystal examined using X-ray diffraction system, D8 DISCOVER (Bruker), confirms that the crystal was cut correctly with 1$^\circ$ accuracy.

The notable feature of the scintillation of a \zwo crystal is its long decay time. A typical scintillation light waveform with $ \gamma $ rays irradiation lasts roughly 100 $ \mu $ seconds, showing a discrete waveform. In prior study, the decay time constant was reported about 10 $ \sim $ 20 $ \mu $ seconds \cite{decay1,decay2,decay3}, although there are differences in each report.

\begin{table}[!h]

\caption{Properties of \zwo crystals}

\label{tb:info}

\centering

\begin{tabular}{c c}\hline

Molar mass(g/mol) & 313.22            \\ \hline

Density (g/cm$^3$) & 7.87            \\ \hline

Reflective index & 2.1 $\sim$ 2.2             \\ \hline      

\end{tabular}

\end{table}%%%End of the table

\begin{table}[!h]

\caption{Angles and lengths of a unit cell of \zwo crystals }

\label{tb:unitcell}

\centering

\begin{tabular}{c c c}\hline

$\alpha$[deg.] & $\beta$[deg.] &$\gamma$[deg.] \\ \hline

90.0000 & 90.6210 & 90.0000            \\ \hline\hline

a[\AA] & b[\AA] & c[\AA]        \\ \hline

4.96060 & 5.71820 & 4.92690     \\ \hline

\end{tabular}

\end{table}%%%End of the table

\subsection{The neutron beam}

\begin{figure}[!h]

\centering\includegraphics[width=5.7in]{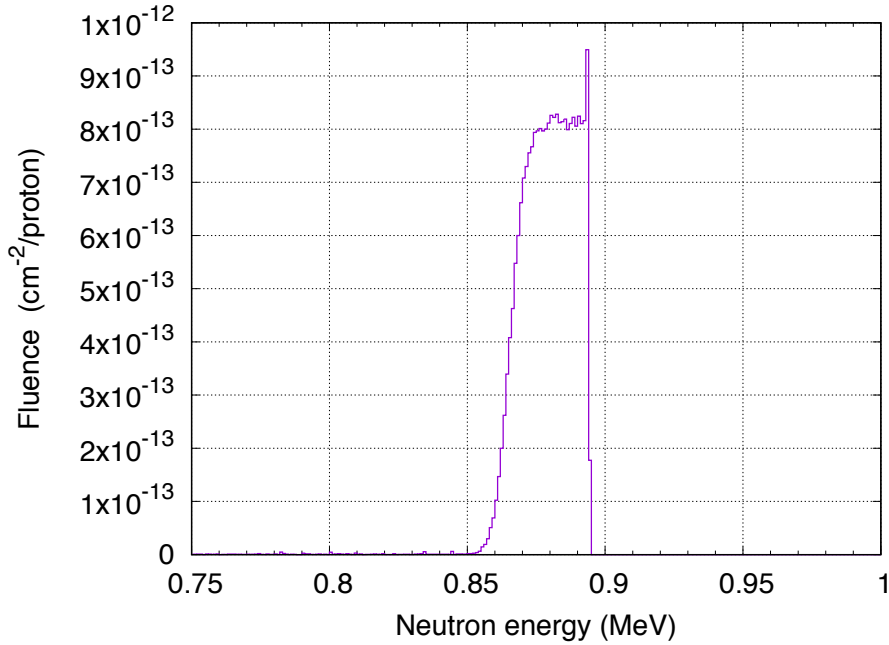}

\caption{The energy spectrum of the irradiated neutron beam. This spectrum was calculated using AIST's neutron generation simulation package based on  MCNP-ANT\cite{beam_sim}.}

\label{fig:beamene}

\end{figure}

To measure the quenching factors for the nuclear recoil of the \zwo crystal, we conducted neutron-scattering experiments using a monochromatic neutron beam by 4 MV Peretron accelerator provided by National Institute of Advanced Industrial Science and Technology on the 26th and 27th of March, 2019.
The pulsed proton beam, which is accelerated at 1.7 MV interacts with a tritium (T) target, generating neutrons in all directions by the T (p, n) reaction. The generated neutron energy is 0.885 MeV on the beam line. Although monochromatic neutrons are generated as a result of the reaction, the energy of the neutron beam has a spectrum with an energy width of about $_{-20}^{+10}$ keV due to the thickness of the tritium target, as shown in Fig. \ref{fig:beamene}. This spectrum was calculated using AIST's neutron generation simulation package based on  MCNP-ANT\cite{beam_sim}. \par
The beam flux is $ \sim $ 310 neutrons/cm$ ^2 $/second at a distance of 100 cm. In this setup, the 2 cm cubed crystal was placed at a distance of 60 cm from the beam source, so the flux per crystal area was $ \sim $ 2,644 neutrons/(2 cm)$ ^2 $/second. \par

\subsection{Experimental setup}

In this study, the A and B surfaces of the \zwo crystal were perpendicularly irradiated with a neutron beam twice to compare the quenching factors of nuclear recoil relative to each surface. We named the beam runs A1 and A2 for the A-surface irradiation and B1 and B2 for the B-surface irradiation. The beam time for each run was 20 minutes.

\begin{figure}[!h]

\centering\includegraphics[width=5in]{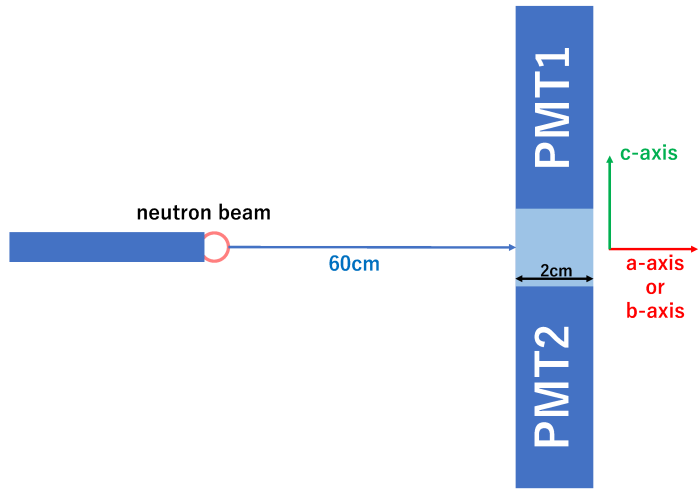}

\caption{A schematics of the experimental setup}

\label{fig:schematics}

\end{figure}

\begin{figure}[!h]

\centering\includegraphics[width=7in]{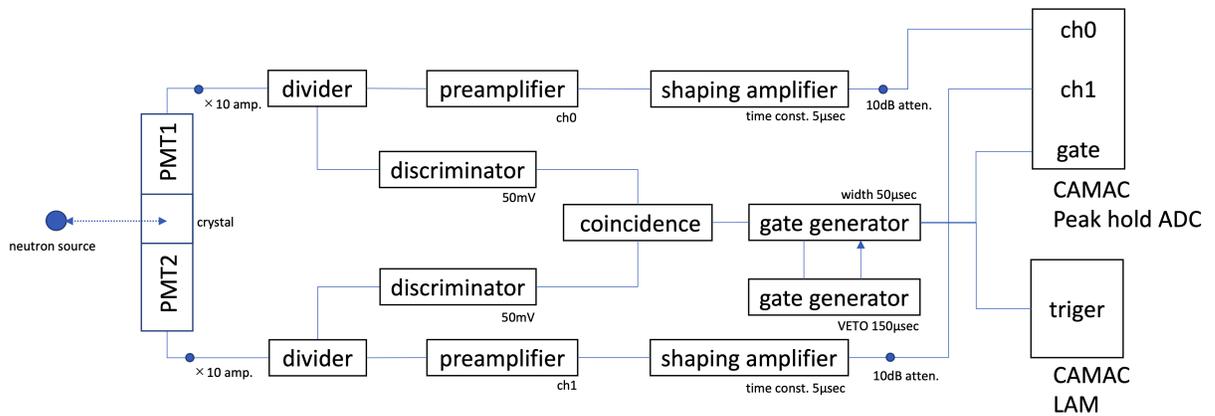}

\caption{A schematic diagram of the measurement circuit}

\label{fig:setup}

\end{figure}

The crystal and PMTs are arranged as shown in Fig. \ref{fig:schematics}, and the distance between the beam source and the crystal was 60 cm. Two PMTs (Hamamatsu H6411) were mounted on the C and C$^\prime$ surfaces with optical grease (Adhesive Materials Group, V-788). PMT1 (attached to the C surface) and PMT2 (attached to the C$^\prime$ surface) were applied at voltages of -2,200 V and -2,140 V, respectively, so that the gains of the PMTs were identical at 1.25$\times$10$^7$. Additionally, Teflon tape was wrapped around the A, A$^\prime$, B, and B$^\prime$ surfaces to improve the collection of scintillation light. The crystal was fixed with a mounter made from Styrofoam$^{\textregistered}$. A schematic diagram of the measurement circuit is given in Fig. \ref{fig:setup}, and a list of modules is given in Table \ref{tb:module}. The output signals from PMT1 and PMT2 were divided and sent to discriminators and their coincidence signal was taken as the overall trigger. The divided analogue signals from the PMTs were pre-amplified, shaped, and sent to peak hold ADCs. The charge-integration time constant of the preamplifier was set at 66$\mu$sec to collect the slow scintillation light from the \zwo crystal. The signal from coincidence was used for the gate of the peak hold ADC for the other PMT signals, and its gate width was set to 50$\mu$ \ seconds. \par

\begin{table}[]
\caption{A list of modules used in measurements}
 \label{tb:module}
\centering

\begin{tabular}{cccc}

\hline

Standard & Module & Model number  \\ \hline
   & PMT & Hamamatsu H6410    \\ %\hline
   & Preamplifier & (Scratch built, op-amp AD817AZ)     \\ %\hline

CAMAC & Crate Controller & TOYO CCNET   \\ %\hline

CAMAC & LAM (Charge ADC) & LeCroy 2249A     \\ %\hline

%CAMAC & Input register & HOSHIN C005   \\ %\hline

CAMAC & Peak hold ADC  & HOSHIN C011     \\ %\hline

CAMAC & TDC & LeCroy 2228A     \\ %\hline

NIM & Discriminator & Kaizuworks KN 240     \\ %\hline

NIM & Shaping amplifier & CLEAR PULSE 4463    \\ %\hline

NIM & Coincidence & Kaizuworks KN 470    \\ %\hline

NIM & Gate generater & Kaizuworks KN 1500    \\ %\hline

\end{tabular}

\end{table}

\subsection{Monte Carlo simulation of nuclear recoils}
\begin{figure}[!h]

\centering\includegraphics[width=5.7in]{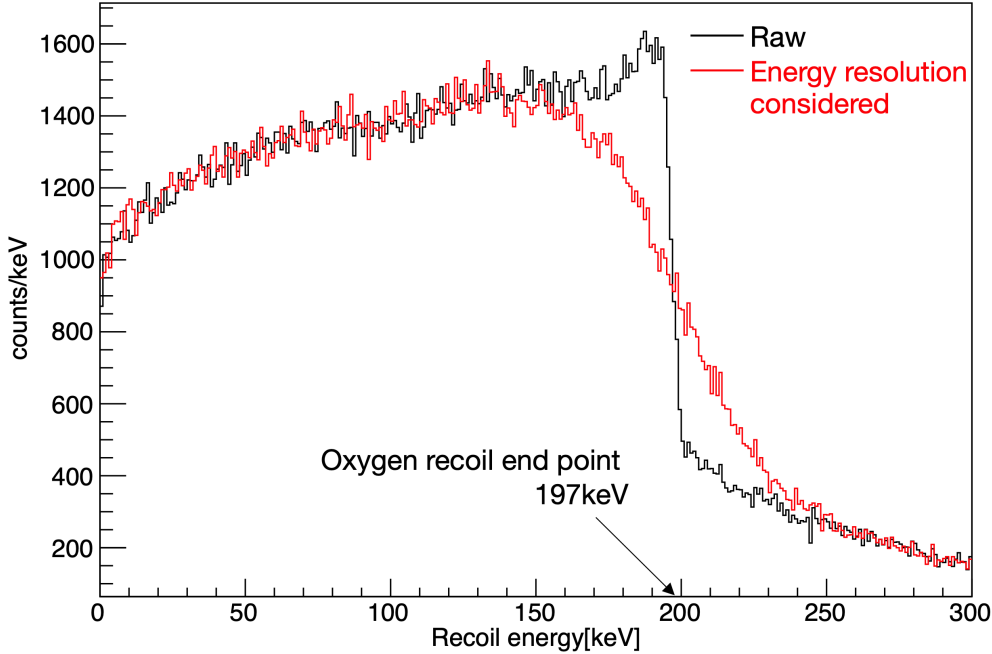}

\caption{A simulation result of oxygen-nucleus recoil energy}

\label{fig:geant}

\end{figure}

The energy spectra obtained by conducting the appropriate experiments on A1, A2, B1, and B2 are superposition of the events scattered according to the angles in the crystal. Therefore, to derive the quenching factor, Monte Carlo simulation of recoil energy spectrum is required.
The simulated oxygen-recoil energy spectrum by Geant 4.9.6 is shown in Fig. \ref{fig:geant}. A histogram drawn with black lines indicates a raw energy deposit by neutrons, and a red line indicates a histogram smeared by Gaussian distribution that considers energy resolution from the statics of the number of photoelectrons. 
The maximum energy deposit of oxygen-nucleus recoil for a 0.885 MeV neutron beam is 0.197 MeV, which corresponds to the events associated with a scattering angle of $\theta = 180^\circ$ and to an edge of the recoil energy $E_R=0.197$ MeV in Fig. \ref{fig:geant}. As the maximum energy deposits of zinc-nucleus recoil and tungsten recoil are 0.059 MeV and 0.022 MeV, these events do not affect the oxygen-recoil edge region and are not taken into account in this simulation.

\newpage

\section{Results}

\begin{figure}[!h]\label{fig:beam}

\centering\includegraphics[width=5.7in]{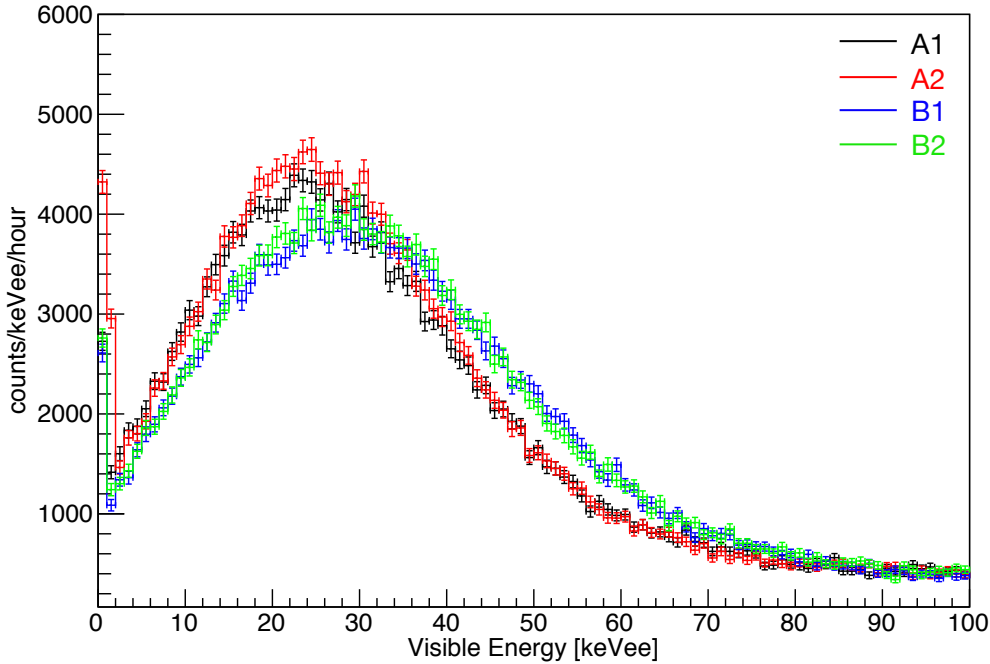}

\caption{Energy spectra obtained from each beam. Black and red histograms are results for the surface A irradiation and blue and green histograms are for the surface B irradiation.}

\end{figure}

We conducted energy calibration using the peak energies of the $ \gamma $ sources \Cd, \Cs, \Am, and \Ba. The results are summarized in Table \ref{tb:gamma} from which it is evident that the scintillation response is linear to gammas in this energy range. The obtained visible energy spectra for the neutron beam runs, which are converted to keVee (electron equivalent) unit according to Table \ref{tb:gamma}, are shown in Fig. \ref{fig:beam}.

\begin{figure}[!h]

\centering\includegraphics[width=5.7in]{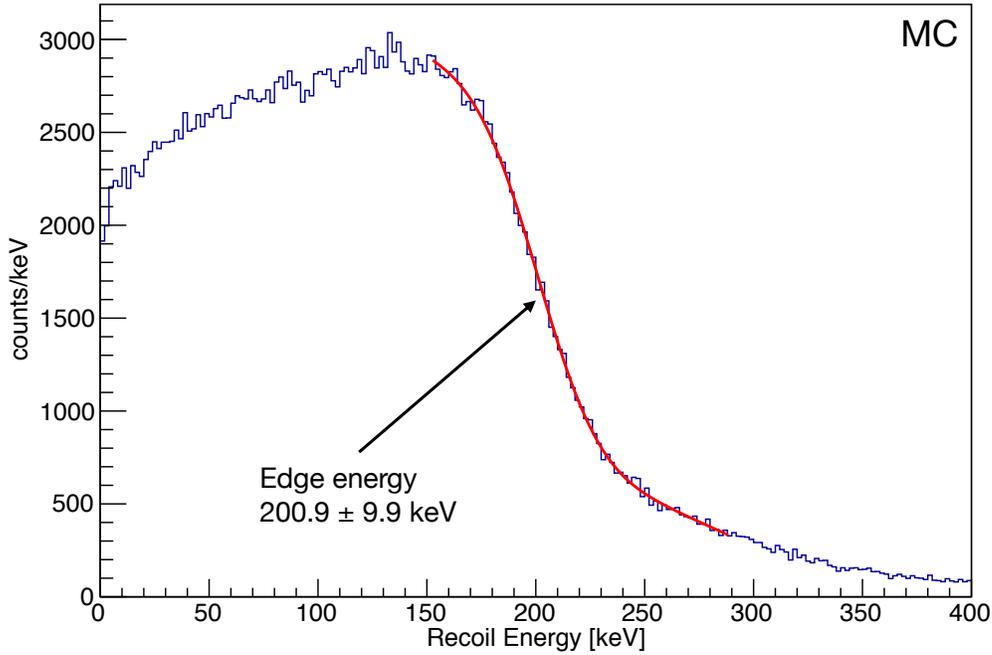}

\caption{Fitting result of the obtained spectrum from Monte Carlo simulation}

\label{fig:geant_fit}

\end{figure}

\begin{figure}[!h]

\centering\includegraphics[width=5.7in]{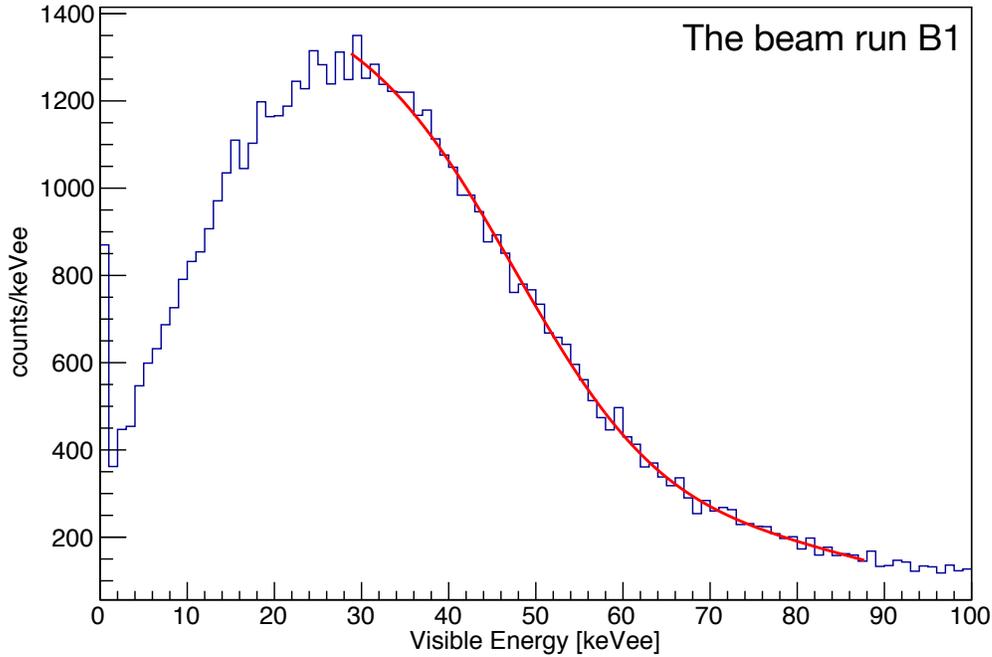}

\caption{Fitting result of the measured spectrum from the beam B1}

\label{fig:beam_fit}

\end{figure}

The difference in light output between the A-surface irradiation and the B-surface irradiation is clearly observed. The quenching factor of each surface is calculated by comparing the visible energy spectrum and the nuclear recoil spectrum (keV) obtained from the Monte Carlo simulation. However, because the recoil energy dependence of the quenching factor is not known here, the quenching factor is derived from the ratio of the ``edge" positions of the spectra of the Monte Carlo simulation to that of the measurement results, by fitting their spectra with ``an error function $+$ a linear function".
The fitting result of the Monte Carlo simulation and the visible energy spectrum of the beam run B1, as an example, are shown in Fig. \ref{fig:geant_fit} and Fig. \ref{fig:beam_fit}, respectively.
The quenching factor obtained comparing their ``edge" positions is considered to refer the energy associated with the oxygen nuclear recoil of $\sim$ 200 keV as indicated in Fig. \ref{fig:geant}.
\par

\begin{table}[!h]

\caption{Peak energies of gamma sources used for the energy calibration and numbers of detected photoelectrons}

\label{tb:gamma}

\centering

\begin{tabular}{ccc}

%\hline

$\gamma$ source & Peak energy (keV) & Number of photoelectrons \\ \hline

$^{109}$Cd & 22.1 & 21.0 $\pm$ 2.25 \\ %\hline

$^{133}$Ba & 81.0 & 74.6 $\pm$ 6.98 \\ %\hline

$^{137}$Cs & 32.2 & 30.2 $\pm$ 3.41 \\ %\hline

$^{241}$Am & 59.5 & 55.5 $\pm$ 5.46 \\ %\hline

\end{tabular}

\end{table}%%%End of the table

\begin{table}[!h]

\caption{Obtained quenching factors from each beam}

\label{tb:beam}

\centering

\begin{tabular}{c c c}

%\hline

Beam & Peak energy (keVee) & Quenching factor \\ \hline  

A1 & 39.8 $\pm$ 3.4  &  0.198 $\pm$ 0.020 \\ 

A2 & 40.2 $\pm$ 3.2  &  0.200 $\pm$ 0.019 \\

B1 & 47.6 $\pm$ 4.6  &  0.237 $\pm$ 0.026 \\ 

B2 & 46.8 $\pm$ 4.7  &  0.233 $\pm$ 0.026 \\

\end{tabular}

\end{table}%%%End of the table

The results are summarized in Table 5. To derive the error for the quenching factors, the systematic error on the energy scale from the energy calibrations (Table 4) and the systematic error on the neutron energy from our experimental configurations, that is the angle error between the surface of the crystal and the beam $\sim$ 2$^\circ$ (although it only contributes less than 1\% of the total error ) were taken into account.\par
As seen from Table \ref{tb:beam}, the obtained quenching factors of the $\sim$ 200 keV oxygen nuclear recoil for the first beams (A1 and B1) and the second beams (A2 and B2) are consistent. Averaging the results from the B1 and B2 beam, the quenching factor of the B surface ($ {\rm QF_B} $) is 0.235 $ \pm $ 0.026, and the quenching factor of the A surface ($ {\rm QF_A} $) is 0.199 $ \pm $ 0.020. The anisotropy value between these two surfaces defined as $ \chi = {| {\rm QF_B-QF_A} |} / {\rm QF_B} $, is $ \chi = $ 0.153.

\newpage

%\newpage
\section{Conclusion}

The two surfaces, which are perpendicular to the a- and b-axis of the \zwo crystal were irradiated with a quasi-monochromatic neutron beam, the central energy of which was 0.885 MeV. The quenching factors of the surfaces for $\sim$ 200 keV oxygen nuclear recoil were measured. 
The measured values were 0.199 $\pm$ 0.020 for the A surface and 0.235 $\pm$ 0.026 for the B surface.
Accordingly, it was confirmed that scintillation yields of the crystal show at least 15.3\% of anisotropy depending on the incident direction of the neutrons, although this result is a superposition of all scattering angle events such as actual dark matter experiments.\par
A preceding work\cite{adamo}  reported the anisotropic response of \zwo scintillator about 30\% for 2$\sim$5 MeV alpha particles. The value of the anisotropy is rather different from this work; however,  it can be explained by the difference of the energy regions and the difference of the radiation sources. While the bulk effect of the crystal is observed with neutron sources, the surface effect of the crystal is observed with alpha sources, therefore it is inferred that another effect was observed in \cite{adamo}.\par
With this measurement, the anisotropic scintillation response of the \zwo crystal to nuclear recoils was confirmed for the first time, and the possibility of the \zwo crystal as a direction-sensitive dark matter detector was verified.

\newpage

\section*{Acknowledgment}

We would like to thank Dr. Tetsuro Matsumoto and Dr. Akihiko Masuda from Radioactivity and Neutron Standards Group, Research Institute for Measurement and Analytical Instrumentation, National Meteorology Institute of Japan, and National Institute of Advanced Industrial Science and Technology for the accelerator operation. We would also like to thank Dr. Shunsuke Kurosawa from New Industry Creation Center, Tohoku University and Yamagata University, as well as Dr. Akihiro Yamaji from Institute for Materials Research, Tohoku University for measuring the orientation of the \zwo crystal. This work was supported by JSPS KAKENHI, with grant numbers 15K13478 and 17H02884.


\begin{thebibliography}{9}

\bibitem{zero} A. K. Drukier et al., Phys.Rev. D33, 3495 (1986)

\bibitem{one} D. N. Spergel, Phys. Rev. D 37, 1353　(1988)

\bibitem{two} J. B. R. Battat et al., Physics Report 662, 1 (2016)

\bibitem{sekiya} H. Sekiya et al., Phys.Lett. B571, 132 (2003)

\bibitem{hisano} J. Hisano et al., Phys.Lett. B690, 311 (2010) 

\bibitem{adamo} F. Capella et al., Eur. Phys. J.C73, 2276 (2013)

\bibitem{fedor} F. A. Danevich et al., Nucl. Instr. Meth. Phys. Res. A544, 553 (2005)

\bibitem{decay1} B. C. Grabmaier, IEEE TNS 31 , 372 (1984)

\bibitem{decay2} I. Holl et al., IEEE TNS35 , 105 (1988)

\bibitem{decay3} L. L. Nagornaya et al., IEEE TNS56, 994 (2009)

\bibitem{beam_sim} MCNP for Accelerator Neutron Target developed by Japan Atomic Energy Agency, https://prodas.jaea.go.jp/PRAD1000




\end{thebibliography}
\end{document}